# The geometry of online conversations and the causal antecedents of conflictual discourse

*An application to climate-related discussions on 4forums.com annotated with ChatGPT*

**Carlo R. M. A. Santagiustina**[1]
Inria, France
Sciences Po médialab, France

**Caterina Cruciani**[2]
Venice School of Management, Ca' Foscari University of Venice, Italy

## Abstract

This article investigates the causal antecedents of conflictual language and the geometry of interaction in online threaded conversations related to climate change. We employ three annotation dimensions, inferred through LLM prompting and averaging, to capture complementary aspects of discursive conflict (such as stance: agreement vs disagreement; tone: attacking vs respectful; and emotional versus factual framing) and use data from a threaded online forum to examine how these dimensions respond to temporal, conversational, and arborescent structural features of discussions. We show that, as suggested by the literature, longer delays between successive posts in a thread are associated with replies that are, on average, more respectful, whereas longer delays relative to the parent post are associated with slightly less disagreement but more emotional (less factual) language. Second, we characterize alignment with the local conversational environment and find strong convergence both toward the average stance, tone and emotional framing of older sibling posts replying to the same parent and toward those of the parent post itself, with parent post effects generally stronger than sibling effects. We further show that early branch-level responses condition these alignment dynamics, such that parent–child stance alignment is amplified or attenuated depending on whether a branch is initiated in agreement or disagreement with the discussion's root message. These influences are largely additive for civility-related dimensions (attacking vs respectful, disagree vs agree), whereas for emotional versus factual framing there is a significant interaction: alignment with the parent's emotionality is amplified when older siblings are similarly aligned. Taken together, these results offer a temporal and structural characterization of how conversation timing and interaction geometry jointly shape the emergence and propagation of conflictual language in online discussions.

[1] E-mail: carlo.santagiustina@inria.fr; carlo.santagiustina@sciencespo.fr

[2] E-mail: cruciani@unive.it

# 1. Introduction

Online platforms have become central arenas for public debate on politically salient and scientifically complex issues, including climate change. While these environments promise broader participation and exposure to diverse viewpoints, a growing body of research shows that they are also prone to conflictual dynamics that can undermine deliberation, discourage participation, and distort the circulation of information and ideas (Bennett & Pfetsch, 2018; Baughan et al., 2021; Altenburger et al., 2024). Understanding how such conflict emerges, evolves, and propagates within online discussions is therefore critical for assessing the democratic implications of social media–mediated communication.

Existing research has shown that conflict in online conversations is not solely a function of issue-based disagreement, but is deeply shaped by how disagreement is expressed. More specifically, linguistic and pragmatic studies distinguish between issue-based disagreement and interpersonal conflict, emphasizing the role of tone, politeness, emotional framing, and respectfulness in determining whether exchanges remain productive or escalate into hostility (Brown & Levinson, 1987; Somasundaran & Wiebe, 2009; Danescu-Niculescu-Mizil et al., 2013). Computational analyses further demonstrate that conversational conflict is dynamic, often evolving from rational argumentation into personal attack through identifiable shifts in sentiment, emotionality, and rhetorical style (Habernal et al., 2018; Wang & Cardie, 2016; Zhang et al., 2018).

More recent work has moved beyond static classification of conflictual language to investigate its temporal and interactional antecedents. Studies of conversational derailment, trolling, and toxic escalation highlight the importance of situational context, interaction structure, and timing, showing that exposure to prior conflict, irregular conversational rhythms, and early interactional cues significantly shape subsequent behavior (Cheng et al., 2017; Roos et al., 2020; D'Costa et al., 2024). These findings suggest that conflict cannot be fully understood by treating messages as independent observations, but instead emerges from structured processes unfolding over time and across interactional positions.

At the same time, much of the existing literature focuses either on pairwise interactions or on aggregate conversational outcomes, leaving the role of local conversational structure underexplored. Threaded discussions are inherently hierarchical: replies are embedded within branching trees, where messages respond not only to a single parent post but also to an evolving local context shaped by earlier sibling replies and by the initial tone of branch-initiating messages. Prior research indicates that early exchanges can establish interactional norms that condition later participation and tone, but systematic evidence on how such norms propagate within discussion branches remains limited (Danescu-Niculescu-Mizil et al., 2013; Zhang et al., 2018).

In this paper, we address this gap by analyzing the geometry of online conversations, which we define as the temporal, sequential, and structural organization of threaded discussions, and its relationship to conflictual language. Using climate change–related discussions from the Internet Argument Corpus, we study how stance (agreement vs disagreement) tone (attacking vs respectful) and emotional versus factual message framing respond to (i) conversational timing, (ii) alignment with parent and sibling posts, and (iii) early branch-level signals. By combining LLM-based annotations with regression models that explicitly account for discussion structure, we provide a compact account of how conflictual discourse emerges and propagates within online conversations.

Our contribution is threefold. First, we show how reply timing and conversational rhythm are systematically associated with shifts in civility, disagreement, and emotional framing. Second, we document strong alignment effects with both parent posts and local sibling contexts, highlighting the importance of meso-level conversational environments. Third, we demonstrate that early branch-level responses moderate subsequent stance alignment, revealing how initial interactional cues shape the trajectory of agreement and disagreement in discussion branches. Together, these findings advance our understanding of conflictual language as a dynamic, structured process and offer new insights into the micro-level and meso-level conversational mechanisms through which online discourse may shape democratic debate.

The remainder of the article is structured as follows. In Section 2 we review existing work on conflictual language, conversational dynamics, and the role of temporal and structural factors in online discussions. In the Methods and Data section (Section 3) we explain the construction of the climate change discussion corpus, the LLM-based annotation procedure, and the representation of discussions as reply trees. In Section 4 we formulate the hypotheses guiding modelling choices and the empirical analysis. In Section 5 we present the results, examining how conversational tone and stance vary with reply timing, parent–child and sibling alignment, and early branch-level signals. Finally, in the Conclusions we discuss the implications of these findings and outline directions for future research.

## 2. Literature review

A growing body of research suggests that conflictual dynamics on social media —manifested through incivility, antagonism, and outrage— pose a challenge to democratic debate by discouraging participation, fragmenting public discourse, and obstructing the circulation of information and ideas across society (Bennett & Pfetsch, 2018; Baughan et al., 2021; Altenburger et al., 2024).

The implications of conflict dynamics on social media suggest that conflictual language is key in understanding the emergence and evolution of conflict, not just looking at the words of conflict but more broadly at the structural features of conflictual language. Conflict in conversations (both online and offline) has been studied from a variety of perspectives, ranging from theoretical, linguistic, or behavioral standpoints. Early studies distinguished thematic conflict, focused on issues, from interpersonal conflict, which targets credibility or identity. Wang and Cardie (2016) modeled disputes in Wikipedia discussions, showing that interpersonal tension often emerges through tone and sentiment rather than explicit contradiction. Similarly, Somasundaran and Wiebe (2009) and Choi and Cardie (2008) identified contrastive discourse cues, negation, and compositional sentiment patterns as early indicators of disagreement.

Linguistic and pragmatic cues play a key role in shaping tone. Drawing on politeness theory (Brown & Levinson, 1987), Somasundaran and Wiebe (2009) and Danescu-Niculescu-Mizil et al. (2013) showed that politeness, hedging, and gratitude correlate with cooperative behavior, while sarcasm and intensifiers mark hostility. Subsequent work (Habernal et al., 2018; Wang & Cardie, 2014) demonstrated that conflict is dynamic, often escalating from rational debate to personal attack through emotional or sarcastic shifts. Modeling these transitions enables systems to detect early signs of breakdown or cooperation.

Canute et al. (2023) define conflicts broadly as ideas, arguments, or attitudes that oppose each other. Similarly, Bolander and Locher (2012) conceptualizes disagreement —the primary vehicle for conflict— as a "discourse move" used to express an opposition to a position, stance, or view. Several authors embrace the idea that conflict is inherently a multidimensional concept, providing different definitions and classifications. D'Costa et al. (2024) draw on traditional management literature to categorize conflict into three types: task conflict (differences in task opinions), relationship conflict (interpersonal disagreements), and process conflict (coordination issues). Zhang et al. (2018) focus on "conversations gone awry", defining conflict through pragmatic devices like forceful questioning or a lack of politeness that signals a departure from civil collaboration.

Conflict is multidimensional and measured via multiple tools (say more), but there is a consensus that conflict is driven by two distinct factors: what is said (the topic) and how it is said (the expression). Looking at the outcome of conflict, D'Costa et al. (2024) find that expression features (sentiment, politeness, turn-taking) are more reliable predictors of destructive conflict than the actual topic being discussed. Zhang et al. (2018) align with this, showing that politeness strategies (or their absence) in the very first exchange of a conversation are the primary indicators of whether it will eventually derail.

Looking at the intensity of conflict, several authors agree on the idea that conflict should be assessed beyond a purely binary (good vs bad) view. Canute et al. (2023) distinguish between agonism (productive conflict between adversaries over interpretation) and antagonism (destructive conflict between enemies focused on delegitimizing the other). Bolander and Locher (2012) distinguish between expected, norm-adhering disagreement (consensual) and face-threatening, aggressive disagreement (conflictual).

Provided that conflict cannot be proxied without embracing a multi-domain perspective, several different strands of literature suggest dimensions that need to be included in the assessment of the temporal evolution of conflict. A first interesting insight comes from the literature on polarization, where Anderson et al. (2019) describe how incivility affects how information and sources are perceived (bias, credibility). Kim and Kim (2019) find that exposure to uncivil disagreement increases polarization via negative emotion, with factuality of supporting information unable to play a mediating role. Thus, while disagreement per se might not necessarily lead to conflict, the politeness with which disagreement is expressed affects the reaction to the content.

Conflictual language can be used both to recognize, classify, and analyse hate speech or conflict that has already been published, and to forecast or identify early warning signs to intervene while a conversation is still salvageable. In the first group, we find several contributions that help operationalize the concept of conflict within social media, like Jahan & Oussalah (2023), who present a systematic review focused on automatic textual hate speech detection. This work outlines the generic pipeline for classification, which involves taking a dataset of existing social media posts and training a model to identify various categories like racism, sexism, or cyberbullying. On a similar note, Anjum & Katarya (2024) focus on online hate speech (OHS) recognition to stop the circulation of toxic messages. The paper reviews traditional machine learning and deep learning approaches used to classify existing content as "hate" or "non-hate."

Moving to papers that explicitly aim at providing a predictive framework for the unfolding of conflictual dynamics, Zhang et al. (2018) focus on identifying linguistic cues, such as politeness strategies and rhetorical prompts, in the first exchange of a conversation to forecast whether it will eventually derail into personal attacks. Canute et al. (2023) emphasize the use of conversational context (number of participants and interaction structure) to detect "early signs" of unproductive disputes. D'Costa et al. (2024) seek to prevent a conflict from escalating by identifying linguistic markers associated with "constructive" versus "destructive" outcomes, framing the research as a way to integrate theoretical approaches into early detection and intervention mechanisms. Imran et al. (2026) explore forecasting derailment specifically in GitHub discussions, identifying derailment points —comments that signal a shift away from productivity toward negativity— which typically occur a median of three comments before the first toxic remark. Ranjith et al. (2025) use time-series forecasting to pinpoint how toxic a conversation will become at a specific future timestamp based on user behavior features.

A substantial body of research shows that online conversations are shaped not only by what users say, but also by when and how they take turns in interaction. Timing, response delays, turn-taking patterns, and participation rhythms are central components of what we call the geometry of conversation—the structured interplay of temporal, sequential, and relational forces through which discursive conflict emerges and evolves.

Regarding conversation rhythm and the time evolution of conflict, Roos al (2020) explores the features of online conversations in comparison with face-to-face conversations. They show that online interactions are less responsive due to lack of synchronicity and less ambiguous due to the absence of subtle social cues than face-to-face talk conversations, and these differences significantly affect how people perceive one another. Slower or uneven response rhythms make participants feel ignored and trigger attributions of disinhibition or hostility, increasing polarization and reducing solidarity. This suggests that temporal gaps between posts—whether long pauses or rapid back-and-forth—are not neutral but shape relational interpretations and tone. Again, D'Costa (2024) shows the relevance of timing in real-life conversations, showing that irregular speaker behavior and timing can be stronger indicators of conflict than topic alone. The relevance of the situational context for the outcome of an online discussion is also able to explain trolling activities in Cheng et al. (2017): they show that trolls are not born but made by a mixture of mood and discussion context, where the tone and emotional versus factual framing set by the parent post or the vicinity to a trolling comment impact significantly the probability to troll across users and not focussing on a specific type. This suggests that the local environment has an effect on the evolution of conflict, which is intrinsically dynamic to the structure of the conversation. Temporal evolution of conflict is crucial in two of the papers already cited regarding the focus on conflict antecedents: Ranjith et al. (2025) explore the time evolution of toxicity in social media database (Twitter and Reddit) and forecasts toxicity using user-behavior features and temporal structure, showing the relevance of temporal antecedent. Finally, Imran (2025) basically implements an idea of early warning signs in the analysis of derailment points on Github, while at the same time LLM-based annotations and predictions. Timing in the conversation has a similar impact on emotional appraisal and the successive evolution of the conversation: as shown in Verma et al. (2025).

Moving on to the dimensions affected by the temporal evolution of conflict, extant research suggests several reasons for tone or content alignment. Setting aside the relevance of social power, which might be less readily applicable to non-recurring online conversations, Xu et al. (2018) explain alignment in form and structure, even in the absence of semantic convergence. This suggests that a more curated post like a parent post in a conversation might have alignment effects on subsequent posts.

## 3. Methods and Data

This study follows a five-step analytical pipeline, summarized in Figure 1. We collected conversational data from online discussions hosted on 4forums.com. We then constructed a topic-specific dataset by filtering the 4forums.com corpus for climate change–related conversations, resulting in over 2,300 posts across more than 60 discussion threads. Next, each reply is annotated (using an LLM) in relation to its parent post along three dimensions capturing stance (agreement vs disagreement) and tone (respectful vs attacking) and emotional versus factual framing. These annotations are then used in regression analyses to test hypotheses about conversational dynamics within discussions, understood as patterns that emerge from both the temporal sequencing of replies and the arborescent (tree-like) structure of threaded discussions.

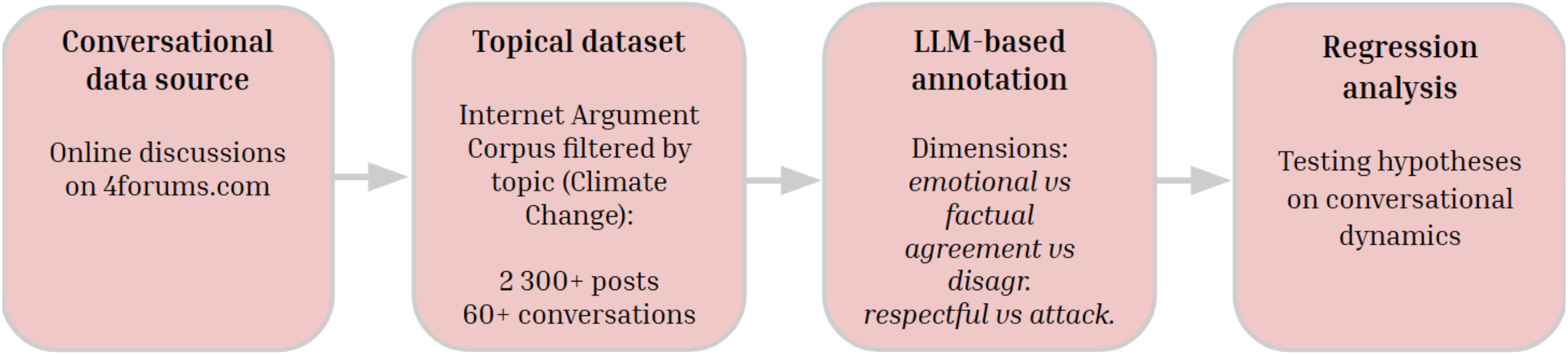

**Figure 1 - Analysis pipeline.**

Further details on each step of this pipeline are provided in the following subsections.

### a. Corpus

The dataset used in this study is drawn from the Internet Argument Corpus (IAC), a large-scale resource developed by the Natural Language and Dialogue Systems (NLDS) Lab at UC Santa Cruz to support computational research on online debates, argumentation, and stance-taking. The first release, IAC v1 (Walker et al., 2012), contains more than three hundred thousand posts across more than ten thousand discussions.

The source platform, 4forums.com, operated as a political debate forum in which users responded to topical prompts (with root posts typically articulating a position or perspective on the issue) spanning domains such as economics, religion, social policy, and environmental issues. The platform supported a threaded reply architecture that enabled branching conversation trees rather than linear exchanges (Abbott et al., 2016). Posts preserve metadata including authorship, timestamps, parent–child reply relationships, quoting behavior, embedded hyperlinks, and formatting features. Although the original forum has since gone offline, the scraped corpus retains richer reply metadata than the platform itself due to manual post-processing efforts during creation of the IAC.

The corpus combines large topical breadth with deep annotation density in politically contentious domains, making it particularly suitable for research on stance modeling, conflict dynamics, and persuasion in online discourse. A feature of this dataset is the presence of human-annotated discourse and argumentation labels. Annotations were collected using crowdsourcing via Amazon Mechanical Turk (Walker et al., 2012).

The Internet Argument Corpus, and particularly the 4forums subset, has been widely used in computational social science, NLP, and argument mining research. Prior works have exploited this corpus for stance detection, interaction modeling, emotional dynamics. For example, Sridhar et al. (2014, 2015) introduced new models for stance classification and demonstrated improvements from incorporating network and relational information over text-only approaches. Deep learning approaches have also leveraged this corpus: Wei-Fan Chen and Ku (2016) developed UTCNN, a neural architecture integrating user and topic embeddings. More recent work explored structural graph embeddings for stance propagation (Pick et al., 2022) and multimodal modeling frameworks combining transformer-based text embeddings with interaction-based conversational structure (Barel et al., 2024). Beyond stance classification, the corpus has also supported research into emotional expression in online political debates (Li & Xiao, 2020), demonstrating its continued relevance across methodological paradigms.

### b. Topical dataset construction

For the purposes of this work, we focus only on discussions related to climate change (i.e., we retain only discussions for which topic_name = "climate change") and use the parent_post_id variable to reconstruct parent–child replies, enabling downstream LLM-based annotation and subsequent analysis. This results in a dataset of 2,375 messages across 62 discussions, containing 2,313 child-parent relations. To visually inspect issues discussed in selected discussions, in Figure 2, we build a word cloud of bigrams (i.e., pairs of consecutive words) extracted from discussion titles. This approach enables the identification of recurring phrase fragments that characterize the selected dataset.

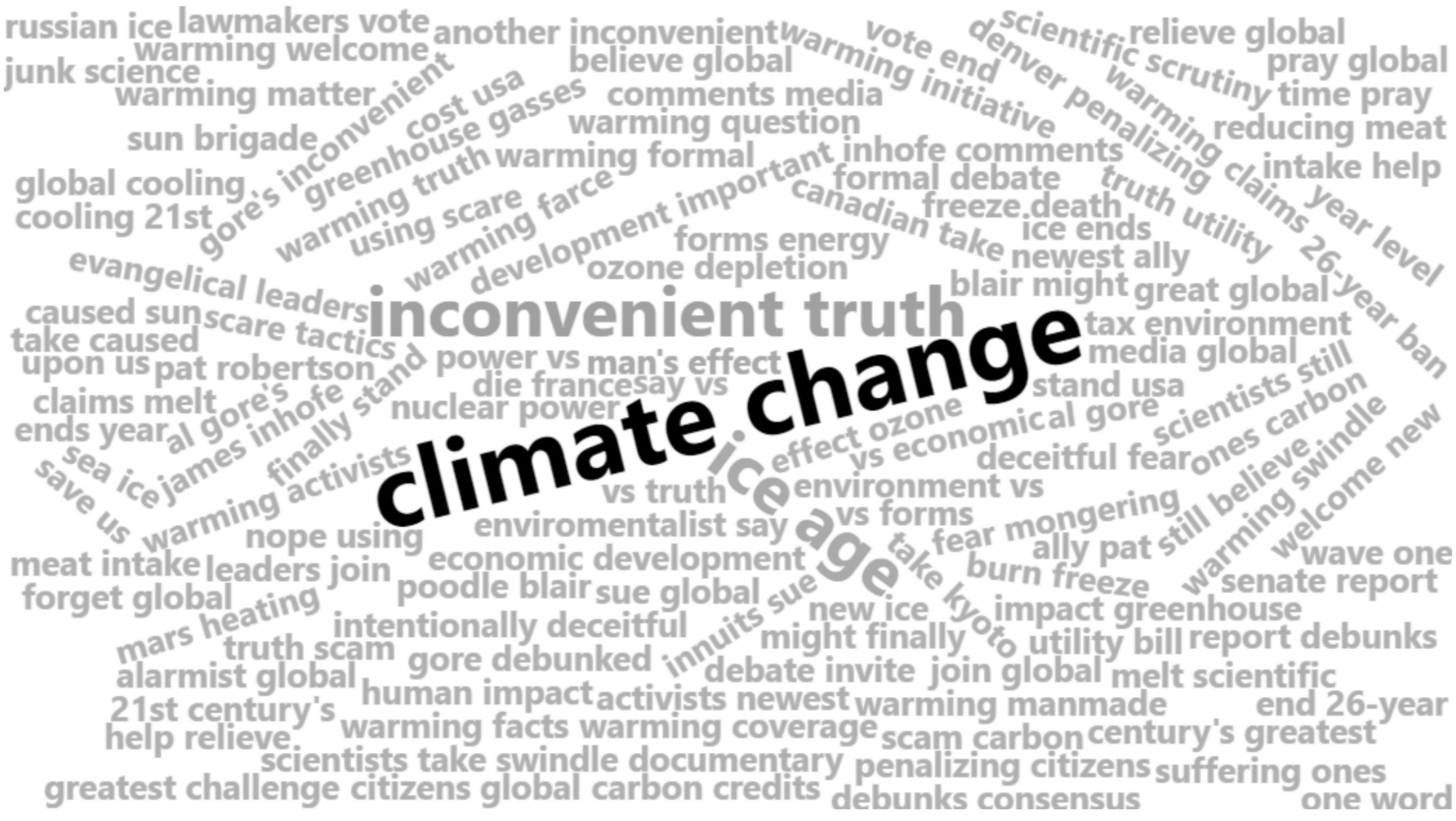

**Figure 2 - Word cloud of bigrams extracted from the titles of selected discussions.**
*Note. The size and darkness of each bigram reflect its relative frequency in the dataset, with larger and darker bigrams representing more frequently occurring phrases*

Online conversations about climate change exemplify the complexity of polarized, science-laden discourse. Research consistently documents that such exchanges are heterogeneous in form and intent, blending factual claims, emotional appeals, and interpersonal hostility (Tyagi et al., 2021). This heterogeneity complicates both interpretive analysis and governance because a single post may simultaneously advance evidence, express affect, and attack an interlocutor —each requiring distinct policy responses (Myers et al., 2023). Emotional framing plays a critical role: anger and fear often amplify polarization, while hope can foster constructive engagement (Liu & Kuang, 2024). Topic modeling studies confirm that climate discourse spans diverse clusters, from scientific evidence to ideological confrontation, reinforcing the need for nuanced annotation frameworks (Böhme & Pfister, 2024).

We hence annotate posts using the original annotation dimensions of the dataset. Unlike the original dataset, which randomly presented the annotator with the parent post and another post from the same branch published earlier, we provide only the parent post to ensure a more stable and reproducible procedure (details are provided in the subsection that follows). The choice of annotating posts in relation to the parent post, to which they reply, is specifically designed to capture stance-taking behavior and tone as a relational characteristic of online conversations rather than as isolated linguistic events, consistent with the design intent of this study.

We decided to reannotate the data using commercial LLMs rather than relying on human-annotated labels because,while manually inspecting annotations before and after averaging, we found that inter-annotator agreement was consistently low for discussions about climate change. The same message was frequently scored at opposite ends of the scale by different non-expert human annotators, for example with some annotators labeling a post as factual and others as emotional, or some judging it respectful and others judging it attacking. This pattern could reflect low-quality MTurk annotations, the effect of the random context provided to annotators, inattentive or random responding by a subset of annotators, differences in how annotators interpreted the dimension definitions, scales and extremes, or genuine semantic ambiguity in the content. Such inconsistencies can also arise from differences in personal perspectives, prior knowledge, or subjective interpretation of tone and stance, particularly when messages contain nuanced arguments, sarcasm, or implicit assumptions. Under any of these explanations, observed major inconsistencies in annotations of climate change related posts can undermine the reliability of mean-based aggregation, especially in relation to our purposes of modeling conflictual discourse dynamics. As averaging these inconsistent annotations produces values close to zero for messages that are possibly “controversial”, and whose annotations may be influenced by annotator’s beliefs (which can be irrational, biased, or otherwise unjustified) in a way that cannot be controlled for (as no user-level characteristics or metadata are available), or that are inherently ambiguous for non-expert annotators (e.g., on complex scientific issues).
When these divergent annotations are combined, the resulting averages can obscure the underlying signal, effectively washing out meaningful information about tone, stance and emotional vs factual framing. This not only reduces the reliability of aggregated scores but also limits the ability of subsequent analyses to detect conflictual discussion patterns, or accurately identify and model discourse dynamics. In cases like ours, the very messages that are most informative for understanding user behavior may be the ones whose signal is most attenuated by conventional non-expert annotations averaging approaches.

For these reasons, and because reannotating manually on online platforms such as Prolific could plausibly produce similarly noisy outcomes and might also encourage annotators to rely on LLMs during the task, we chose to switch directly to LLM-based annotations. While LLM outputs can also exhibit biases, in recent years, large encoder-decoder transformer models (Vaswani et al., 2017) have been increasingly used as automated annotators (Tan et al., 2024), fact-checkers (Leippold et al., 2025), and even as bridges in deliberation processes (Ma et al., 2025). Their appeal lies in their (alleged) strong zero- or few-shot performance on annotation tasks (Le Mens & Gallego, 2025) using natural-language instructions, as well as the ability of a single LLM to handle multiple annotation tasks without task-specific training. Other advantages of delegating this inference to LLMs include: (i) higher annotations consistency across the corpus, (ii) labels that are immediately actionable for machine-based content flagging and moderation, (iii) the possibility of deploying them in online environments where these metrics are computed in real time to provide feedback and nudges to users, and (iv) improved scalability and reproducibility through minimal prompts and versioned LLM models; the drawbacks include: (i) sensitivity to model choice, parameters and prompting strategy, (ii) the potential amplification of social and topical biases, (iii) limited transparency and interpretability relative to carefully trained human expert judgments, which, in this work motivated mitigation steps such as minimal prompt design, annotated dimensions’ definitions identical to those provided to humans and embedded “as is” in the prompt.

## c. LLM-based annotations and prompting details

To annotate conversational stance and tone, we rely on LLM-based annotations designed to closely replicate the original human annotation protocol used in the Internet Argument Corpus. The system prompt used to elicit annotations from the model is constructed directly from the dimension definitions, descriptions, and numeric scales reported in the original MTurk-based annotation study (Walker, 2012). We deliberately keep the prompt as simple as possible. As illustrated in Figure 3, dimension-level definitions and scales are embedded in the prompt with minimal instructions.

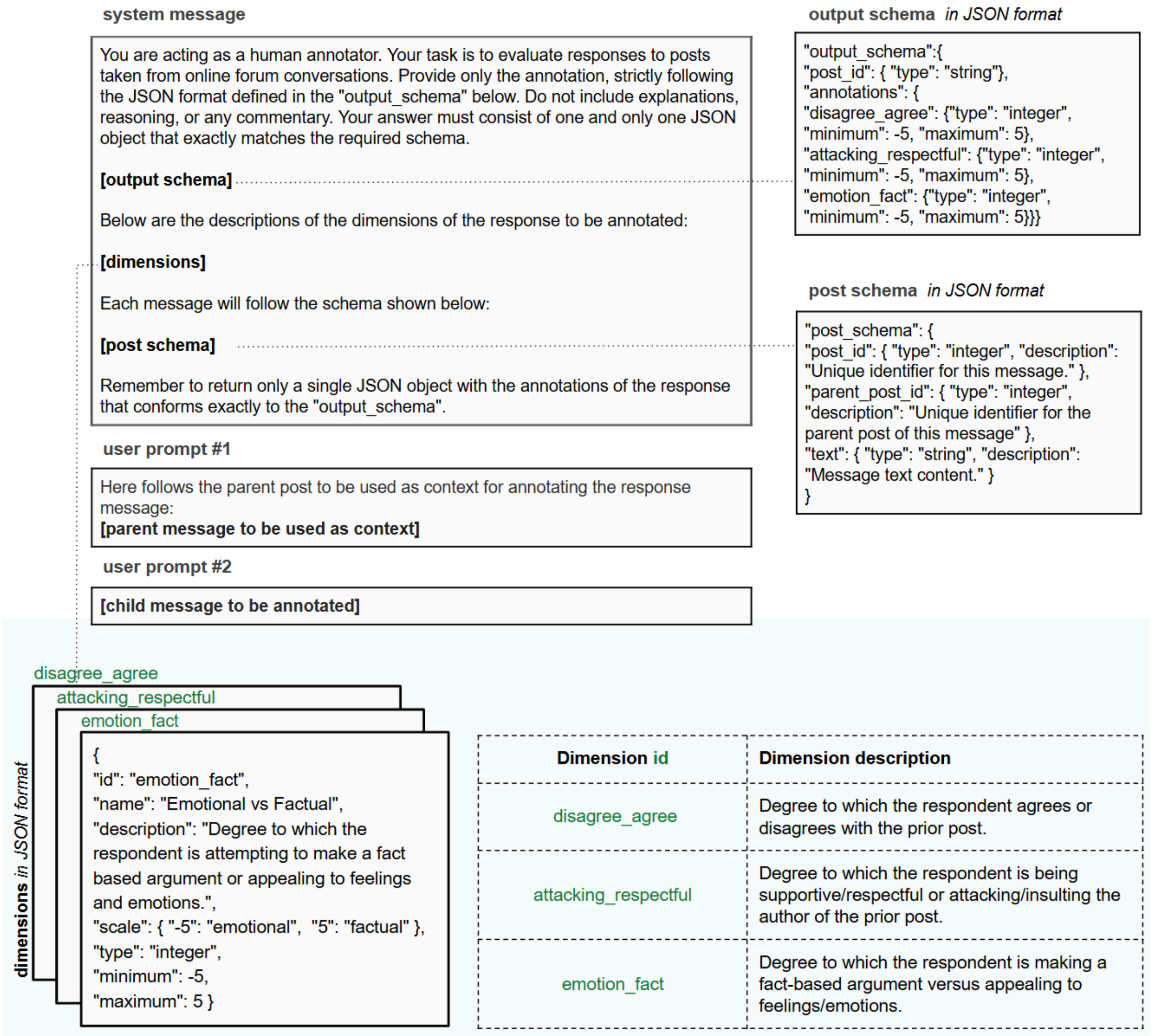

**Figure 3 - LLM-based annotation prompt, schemas and dimensions' descriptions.**
*Note. Structure of the annotation prompt used in this study. The model is instructed to act as a human annotator and to score each reply (child post) in relation to its parent post. The prompt embeds the annotation dimensions' definitions and numeric scales from the original annotation task, and requires output in a fixed JSON format. Only the three dimensions used in the analysis that follow are shown.*

This design choice minimizes the introduction of prompt-induced interpretive biases. As detailed in Figure 3, the annotation prompt consists of a system message that: (1) frames the model as a human annotator, (2) instructs it to interpret each reply (child post) in relation to its parent post, (3) requires the output to consist of a single JSON object strictly conforming to a predefined output_schema, and (4) embeds the item-by-item dimension definitions and numeric ranges. The model is explicitly instructed not to produce explanations, reasoning, or any text outside the JSON object. Enforcing this constraint reduces post-processing and ensures structural parity with the original human-coded dataset.

Each annotation instance includes two user-provided inputs: the parent post, which serves as the sole conversational context, and the child post, which is annotated relative to that parent. Each parent–child pair is processed independently to prevent contextual contamination across annotations. Although the original dataset includes additional annotation dimensions, in this study we focus on three dimensions central to conversational dynamics: disagree vs agree, attacking vs respectful, and emotional vs factual. Figure 3 summarizes the full prompt structure, output schema, and dimension definitions, while Figure 4 provides a concrete example of the input messages and the corresponding model-generated annotation.

We perform annotations using OpenAI's GPT-5.1-mini model (Chat Completions endpoint) accessed via API[3]. This model offers a favorable balance of reasoning capability, computational efficiency, and cost, making it suitable for our annotation task. Since it is not possible to set the temperature to zero for this model, we generate four independent annotations for each observation (i.e., parent–child pair) and average the resulting scores. In addition, we set the effort level to high and verbosity to low to encourage careful inference while strictly constraining output format.
Pairwise correlations between the three post-level dimensions, computed at the observation level after averaging across the four annotation replications, show meaningful but differentiated relationships. The strongest association is observed between attacking vs. respectful and disagree vs. agree (Spearman's $\rho = 0.56$), indicating that more disagreeing replies also tend to adopt a more attacking tone. A moderately strong correlation is found between attacking vs. respectful and emotional vs. factual ($\rho = 0.48$), suggesting that emotionally framed messages are more likely to be confrontational than factual ones. In contrast, the correlation between disagree vs. agree and emotional vs. factual is close to zero ($\rho = 0.02$), implying that, in our dataset, disagreement is largely orthogonal to whether a reply is expressed in emotional or factual terms.

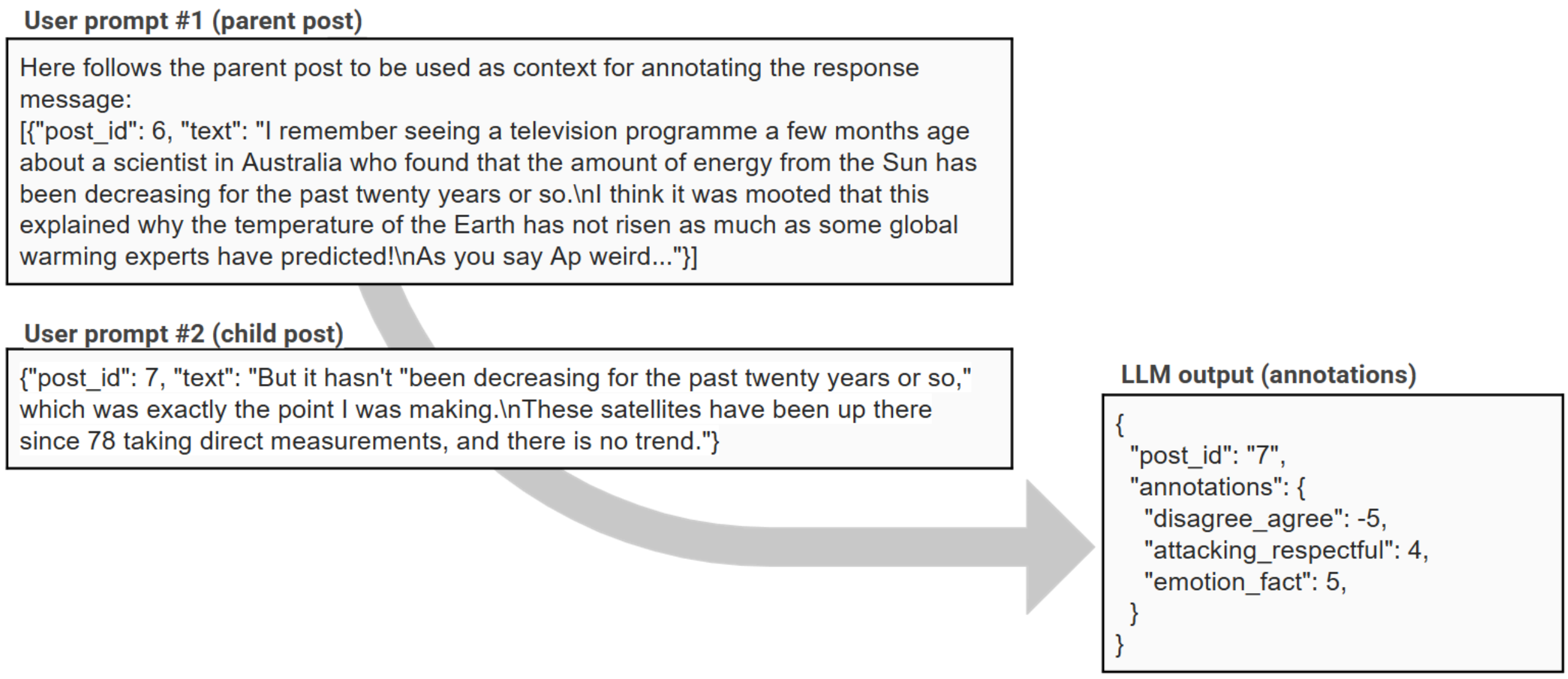

**Figure 4 - Example of inputs and output of LLM-based annotation task.**
*Note. Both parent and child posts are provided to the LLM in the same prompt. The child post is annotated in relation to the parent post. Each observation is annotated through a separate instance to ensure that annotations do not influence one another. As in the original human annotations, more dimensions were annotated than those shown here; we present only the three dimensions used in this work.*

[3] see https://platform.openai.com/docs/models/gpt-5-mini

Table 1 reports intra-annotator agreement across the three annotation dimensions, computed over four independent annotations for each parent–child pair. Overall, agreement levels are rather high across all dimensions, indicating that the annotation procedure yields stable and internally consistent judgments.
Agreement is highest for the disagree vs agree dimension, suggesting that conversational stance relative to the parent post is identified reliably. Agreement is slightly lower for the emotional vs. factual and attacking vs respectful dimensions, which likely reflects the greater interpretive ambiguity associated with distinguishing stance and affective language vs fact-based argumentation.

**Table 1 - "Intra"-annotator agreement of LLM-based annotations across four replications.**

| Dimension | N. obs per rater | Raters | Krippendorff alpha | Fleiss Kappa | MAPD mean | Exact agreement | % within +- 1 | Mean range | Mean SD |
|---|---|---|---|---|---|---|---|---|---|
| disagree vs agree | 2313 | 4 | 0.874 | 0.542 | 0.779 | 0.382 | 0.760 | 1.428 | 0.706 |
| attacking vs respectful | | | 0.885 | 0.471 | 0.894 | 0.280 | 0.742 | 1.623 | 0.799 |
| emotional vs factual | | | 0.921 | 0.471 | 0.787 | 0.290 | 0.755 | 1.433 | 0.701 |

*Note. Agreement metrics are calculated for each dimension across four independent replications with the same model. Metrics include Krippendorff's alpha, Fleiss' kappa, mean absolute pairwise deviation (MAPD), exact agreement, agreement within ±1, mean range, and standard deviation. "N. obs per rater" indicates the number of observations annotated per replication.*

## d. Discussion graphs and conversational dynamics

Online discussions in our dataset are organized as reply trees and can be formally represented as directed, rooted graphs. Each node corresponds to an individual message, and each directed edge encodes a reply relation from a child message to its parent. This representation captures both the temporal ordering of messages and the hierarchical structure of interaction inherent in threaded online discussions (see an example of discussion from our dataset in Figure 5).

Each discussion graph contains a single root message (square node in black in Figure 5) corresponding to the initial post that initiates the discussion. Messages that directly reply to the root define the roots of distinct branches (triangle nodes in Figure 5), each representing an independent conversational trajectory responding to the same discussion initiating post. Subsequent replies extend these branches to greater depths, producing a set of nested interaction sequences within a discussion. Messages that share the same parent are defined as siblings, and their relative ordering reflects the accumulation of local conversational context over time.
As previously explained, the position of a message within a discussion graph has important implications for its conversational dynamics. Branch-root messages (i.e., triangles in Figure 5) are expected to play a particularly salient role, as they establish early signals of stance and tone that may shape subsequent participation within a branch. Deeper messages are embedded within increasingly constrained local contexts, as they respond not only to a parent post but also to the evolving conversational climate created by earlier sibling replies. As a result, conversational behavior is shaped jointly by dyadic interactions (parent–child relations) and by meso-level contextual effects operating within branches and sibling relations.

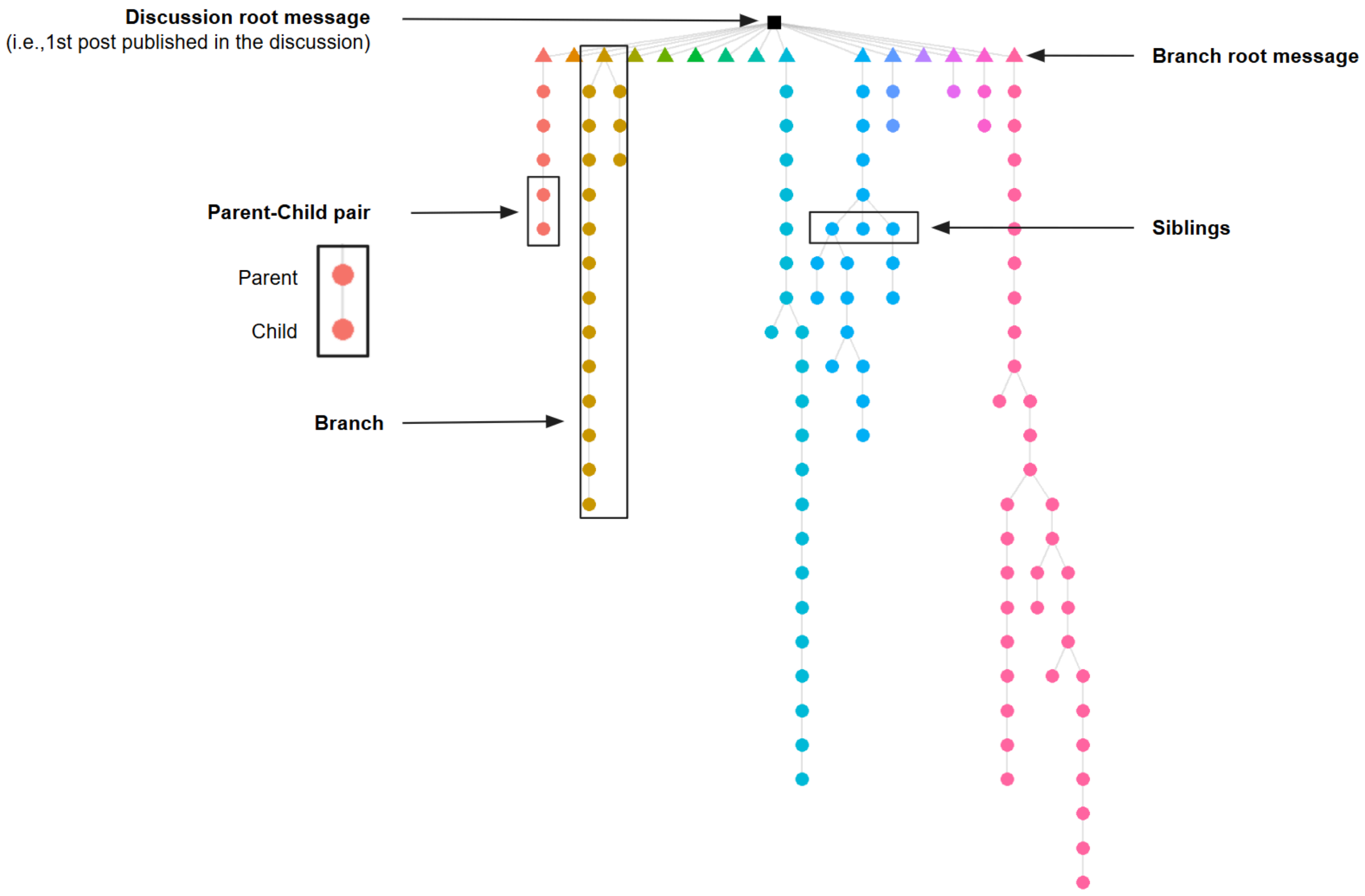

**Figure 5 - Example of graph representation of a discussion (Discussion No. 234).**
*Note. Nodes represent individual messages, and directed edges indicate reply (parent-child) relations. Distinct colors encode branches (all child nodes of a reply to the root message). Node shapes denote message type: squares represent the root message, triangles△ represent the roots of branches (i.e., replies to the root message), and circles○ represent all other messages.*

Several structural and temporal properties of discussion graphs are therefore central to our analysis. First, reply depth captures progression within a conversational trajectory and may be associated with shifts in tone, escalation, or attenuation of conflict. Second, sibling context operationalizes local social influence, as participants may align their responses with the prevailing stance, tone or emotional framing established by earlier replies to the same parent. Third, temporal features, such as delays between consecutive posts and delays relative to the parent post, can capture conversational rhythm and pacing, which have been shown to affect affective intensity and engagement in online interaction (Tsvetkova et al. (2016), Habernal et al. (2018), Zhang et al. (2018)). Finally, early branch-level signals, particularly those introduced by the first reply to the root, may establish local norms that condition how agreement, disagreement, and emotional framing propagate through the remainder of the branch (Danescu-Niculescu-Mizil et al. (2013), Cheng et al. (2017), Altenburger et al. (2024)). This discussion-graph-based perspective allows us to move beyond treating messages as independent observations and instead model conversational dynamics as processes unfolding over time and threaded structures.

In the section that follows we use this framework to derive and test hypotheses about how stance and tone evolve as a function of temporal spacing, parent–child alignment, sibling influence, and branch-level contextual moderation.

## 4. Hypotheses

Table 2 summarizes the hypotheses tested in this study and the strands of literature that motivate them. Rather than treating each hypothesis in isolation, we organize them around three macro-level mechanisms through which conflictual language is expected to emerge and propagate in threaded online conversations: temporal dynamics, local alignment and social influence, and early branch-level norm setting.

**Temporal dynamics (H1.a–H1.c).**
The first group of hypotheses concerns the role of conversational timing and rhythm in shaping the tone of replies. Prior research shows that response delays, pacing, and temporal irregularities affect affective intensity, perceived hostility, and escalation in online interaction (Roos et al., 2020; Cheng et al., 2017). Studies of conflict escalation and derailment further suggest that rapid back-and-forth exchanges are associated with higher emotional arousal and antagonism, whereas longer temporal gaps can attenuate impulsive reactions and reduce incivility (Habernal et al., 2018; Tsvetkova et al., 2016; Ranjith et al., 2025). Together, this literature motivates hypotheses predicting that longer delays since the previous or parent post are associated with more respectful replies (H1.a), systematic shifts in the agree–disagree balance (H1.b), and lower emotional intensity relative to factual framing (H1.c).

**Local alignment and social influence (H1.d–H1.f).**
The second group of hypotheses focuses on alignment with the local conversational environment, operationalized through parent–child and sibling relationships in discussion trees. A substantial body of work shows that exposure to nearby discourse shapes subsequent stance and tone through imitation, normalization, and contextual framing effects (Cheng et al., 2017; Altenburger et al., 2024). Research on conversational derailment and linguistic accommodation further indicates that replies tend to mirror the politeness, disagreement style, and rhetorical framing of preceding messages (Zhang et al., 2018; Danescu-Niculescu-Mizil et al., 2013), while even low-level features of earlier contributions can induce alignment in later responses (Xu et al., 2018). In addition, work on emotional contagion shows that affective framing propagates locally within interaction networks, shaping the emotional tone of subsequent contributions  (Ferrara and Yang ,2015). These findings motivate hypotheses predicting that new posts align with the tone of their parent post and older siblings along dimensions of respectfulness (H1.d), agreement (H1.e), and emotional versus factual framing (H1.f).

**Early branch-level norm setting (branch-root moderation hypothesis H1.g).**
Finally, the last hypothesis addresses the role of early branch-level signals, particularly those introduced by the first reply to a discussion root. Prior studies show that the initial exchange of a conversation plays a disproportionate role in shaping its subsequent trajectory, with early disagreement —or antagonism— strongly predicting escalation or cooperation downstream (Zhang et al., 2018; Danescu-Niculescu-Mizil et al., 2013). Exposure-based models of online behavior further suggest that early local context conditions how later participants interpret and respond to subsequent messages (Cheng et al., 2017; Altenburger et al., 2024). This literature motivates the hypothesis that parent–child alignment dynamics are moderated by the stance of the branch-root message, such that agreement and disagreement propagate differently depending on whether a branch is initiated in alignment or opposition to the discussion root.

Together, these hypotheses translate existing insights on conversational timing, social influence, and norm formation into a unified discussion-graph-based framework, allowing us to model conflictual discourse as a process unfolding jointly over time and interactional discussion thread structure.

**Table 2 - Research question, response metrics and supporting literature**

| Research question | Response metric | H1 description (expected causal direction) | References |
|---|---|---|---|
| How does time (since the previous / parent post in the same discussion affect the tone of a new post? | attacking vs respectful | **H1.a** Longer time since the previous / parent post is associated with more respectful replies (longer gaps → more respectful) | Longer delays since the previous or parent post are expected to be associated with more respectful replies, as prior work shows that conversational pacing and temporal distance can attenuate impulsive reactions and reduce affective escalation in online interaction (Roos et al., 2020; Cheng et al., 2017; Habernal et al., 2018) |
| | disagree vs agree | **H1.b** Time since the previous / parent post affects agree-disagree balance | The balance between agreement and disagreement is expected to vary with temporal spacing between posts, given evidence that interaction rhythm and conversational sequencing shape how oppositional stances are expressed and sustained over time (Tsvetkova et al., 2016; Tan et al., 2016; Roos et al., 2020) |
| | emotion vs fact | **H1.c** Time since the previous / parent post is associated with less emotional reactions | Longer delays since the previous or parent post are hypothesized to be associated with less emotional and more factual replies, consistent with findings that rapid exchanges amplify affective intensity while slower conversational rhythms dampen emotional escalation (Cheng et al., 2017; Roos et al., 2020; Altenburger et al., 2024) |
| How closely do new posts align with the tone of the parent post / older siblings (i.e., children of the same parent post)? | attacking vs respectful | **H1.d** More respectful parent post / older siblings → more respectful new post | Replies are expected to align with the respectfulness of parent posts and older siblings, as exposure to nearby discourse has been shown to normalize and propagate civility or incivility through local imitation and contextual framing effects (Zhang et al., 2018; Cheng et al., 2017; Altenburger et al., 2024) |
| | disagree vs agree | **H1.e** More agreeing parent post / older siblings → more agreeing new post | Agreement and disagreement in replies are expected to align with the stance of parent posts and older siblings, reflecting prior evidence that local conversational context |
| | | | and early exposure shape subsequent stance-taking and oppositional dynamics (Cheng et al., 2017; Zhang et al., 2018; Altenburger et al., 2024) |
| | emotion vs fact | **H1.f** More factual parent post / older siblings → more factual new post. | Replies are expected to align with the emotional or factual framing of parent posts and older siblings, consistent with research on emotional contagion and affective spillovers in networked interactions (Ferrara & Yang, 2015; Roos et al., 2020; Altenburger et al., 2024) |
| Do early branch responses, in particular first replies to root posts, moderate parent–child agreement alignment dynamics? | disagree vs agree | **H1.g** Parent–child agreement alignment is moderated by the stance of the branch-root message, such that alignment is stronger in branches initiated by agreement with the discussion root than in branches initiated by disagreement. | Parent–child agreement alignment is expected to be moderated by the stance of the branch-root message, as prior work shows that early interactional signals disproportionately shape local norms and condition the trajectory of subsequent conversational exchanges (Zhang et al., 2018; Danescu-Niculescu-Mizil et al., 2013; Cheng et al., 2017) |

## 5. Results

In this section, we examine how conversational characteristics of messages vary as a function of the temporal and structural organization of discussions. Using post-level annotations of stance (agreement vs disagreement), interpersonal tone (attacking vs respectful), and emotional framing (emotional vs factual), that capture different aspects of the discursive dynamics of conflict in conversations, we test the hypotheses outlined in the preceding section through a series of linear regression models. As anticipated, these three dimensions incorporate information that the literature has established as relevant in the evolution of conflictual language.

Despite the simplicity of the proposed models, and relations between posts in the same discussion, for the considered dimensions, they allow us to infer if and how reply timing, parent–child relationships, and local conversational context shape discursive dynamics within threaded discussions related to climate change.

First, in Subsection a., we focus on temporal aspects, asking how delays between posts, as well as between replies and their parent posts, are associated with shifts in tone and stance.

Then, in Subsection b., we turn to structural alignment effects, assessing the extent to which posts echo the tone and stance of their parent messages and of earlier sibling replies within the same branch.

Finally, in Subsection c., we examine how these influences interact, and whether early branch-level responses condition the transmission of tone and stance as discussions unfold. Taken together, these results provide a systematic characterization of how conflict-related dynamics emerge from the timing and tree structure of online discussions.

***Remark:*** *We are aware that some of the relationships we investigate may exhibit non-linearities; however, for reasons of simplicity and parsimony, we limit the present analysis to linear specifications. In addition, because individual-level demographic variables and user characteristics are not available in the data, we are unable to control for their potential effects. As a result, our models may be subject to omitted-variable bias, a limitation that should be considered when interpreting the results. Nevertheless, we believe that the present work offers valuable insights into the temporal and structural mechanisms through which conflictual language emerges and propagates in online threaded discussions, and provides a starting point for future research that can build on richer user-level data and more flexible modeling frameworks.*

### a. Hypotheses 1a-1.c: the effects of replying delays and conversation rhythm

*Question: How does the time elapsed since the posting time of the previous post in a discussion affect the tone of a reply?*

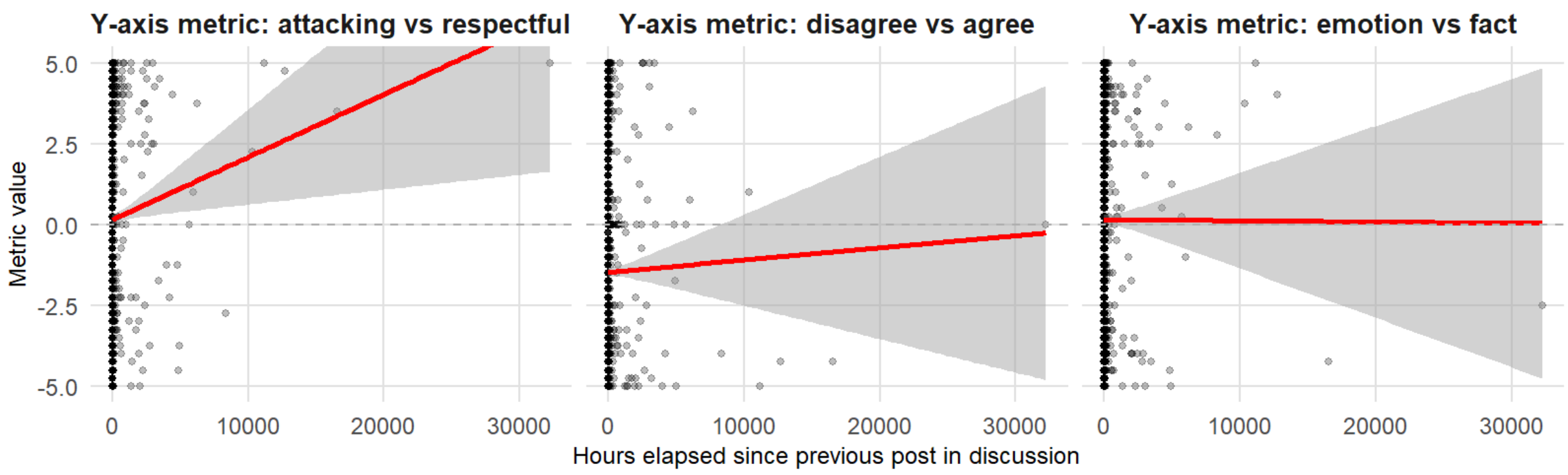

**Figure 6 - Scatter plots of post-level metric values against the time elapsed since the publication of the previous post in the same discussion, with linear regression lines (in red) and associated 95% confidence intervals (in gray).**

For each metric *m*, we estimate a separate linear regression model where the value of the metric for a post *i* is related to the time that has elapsed (in hours) since the previous post in the same discussion. This to assess whether the metric tends to increase or decrease as the time gap between posts becomes longer. The model is specified as follows:

$$y_{i,m} = \alpha_m + \beta_m \, \Delta t_i^{\text{prev}} + \varepsilon_{i,m}$$

We use discussion-level cluster-robust standard errors, which are consistent under heteroskedasticity and correlation of errors within discussions:

$$\widehat{\text{Var}}(\hat{\beta}_m) = (X^\top X)^{-1} \left( \sum_d X_d^\top \hat{\varepsilon}_d \hat{\varepsilon}_d^\top X_d \right) (X^\top X)^{-1}$$

**Table 3 - Linear regression estimates of post-level metrics as a function of the time elapsed since the previous post (in hours) in the same discussion.**

| response metric | term | coeff estimate | std.error | p.value | |
|---|---|---|---|---|---|
| attacking vs respectful | intercept α | 0.14800 | 0.20900 | 0.47900 | |
| | slope β | 0.00019 | 0.00005 | 0.00011 | *** |
| disagree vs agree | intercept α | -1.47000 | 0.14300 | < 0.00001 | *** |
| | slope β | 0.00004 | 0.00004 | 0.33300 | |
| emotion vs fact | intercept α | 0.15900 | 0.17900 | 0.37400 | |
| | slope β | 0.00000 | 0.00008 | 0.96200 | |

*Sign.level: < 0.001 ~ "***" < 0.01 ~ "**", < 0.05 ~ "*", < 0.1 ~ "†"*

*Note. For each metric, the table reports the intercept and the coefficient on time since the previous post (in hours), along with cluster-robust standard errors (by discussion), p-values, and significance levels.*

As shown in Table 3, for the attacking vs respectful metric, the slope on time since the previous post is positive and highly significant ($p < .0001$), indicating that as more time passes between posts, messages become on average slightly more respectful. The only significant intercept is for the disagree vs agree metric, which is negative and statistically different from zero. This implies that, on average, messages in our corpus tend to be more disagreeing than agreeing.

*Question: How does the time elapsed since the posting time of the parent post in a discussion affect the "tone" of a post in the same conversation?*

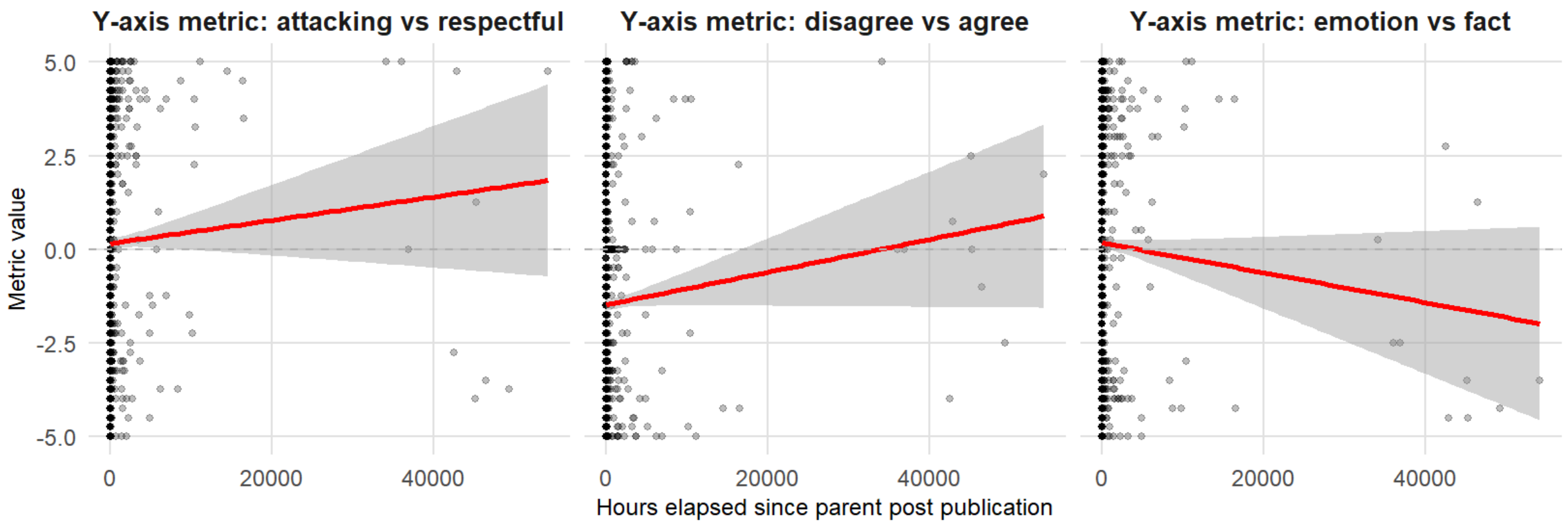

**Figure 7 - Scatter plots of post-level metric values against the time elapsed since the publication of the parent post, with linear regression lines (in red) and associated 95% confidence intervals (in gray).**

For each metric, we estimate a separate linear regression model, whose specification is detailed here below, where the value of the metric for a post is related to the time that has elapsed (in hours) since the publication of the parent post in the same discussion. This to assess whether the metric tends to increase or decrease as the time gap with the parent post becomes longer. The model is specified as follows:

$$y_{i,m} = \alpha_m + \beta_m \, \Delta t_i^{\text{parent}} + \varepsilon_{i,m}$$

As before, we use discussion-level cluster-robust standard errors.

**Table 4 - Linear regression estimates of post-level metrics as a function of the time elapsed since the parent post (in hours).**

| response metric | term | coeff estimate | std.error | p.value | |
|---|---|---|---|---|---|
| attacking vs respectful | intercept α | 0.15800 | 0.21000 | 0.45000 | |
| | slope β | 0.00003 | 0.00002 | 0.19300 | |
| disagree vs agree | intercept α | -1.48000 | 0.13900 | < 0.00001 | *** |
| | slope β | 0.00004 | 0.00001 | < 0.00001 | *** |
| emotion vs fact | intercept α | 0.17300 | 0.18100 | 0.33900 | |
| | slope β | -0.00004 | 0.00001 | 0.00245 | ** |

*Sign.level: < 0.001 ~ "***" < 0.01 ~ "**", < 0.05 ~ "*", < 0.1 ~ "†"*

*Note. For each metric, the table reports the intercept and the coefficient on time since the parent post (in hours), along with cluster-robust standard errors (by discussion), p-values, and significance levels.*

Table 4 highlights that for the disagree vs agree metric, replies are on average more disagreeing than agreeing overall, but this tendency weakens as the time since the parent post increases.

For the emotion vs fact metric, replies become slightly more emotional (less factual) as the delay since the parent grows.

### b. Hypotheses 1.d-1.f: Alignment with local conversation tone and emotional vs factual framing

*Question: How closely do posts align with the tone and framing of older siblings (children/leafs of the same parent post) within a discussion?*

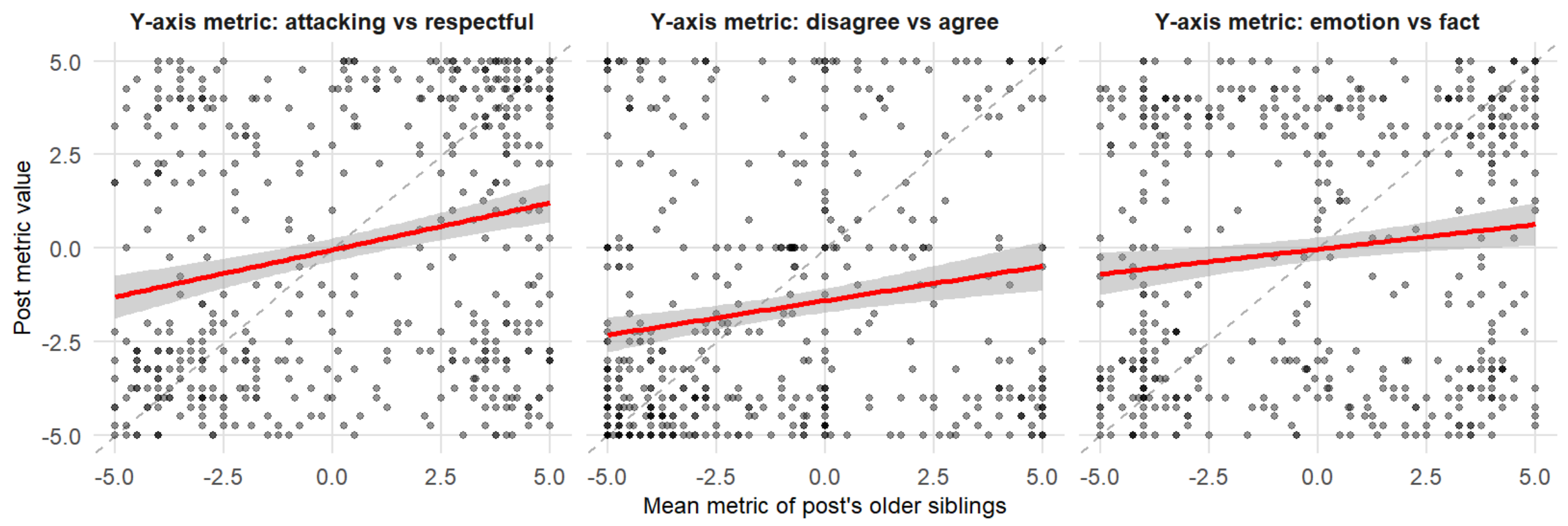

**Figure 3 - Scatter plots of post-level metric values as a function of mean values of the corresponding metric for older siblings of the post, with linear regression lines (in red) and associated 95% confidence intervals (in gray).**

For each metric, we estimate a separate linear regression where the value of the metric for a post is related to the mean value of the corresponding metric for older siblings (childs of the same parent post) within the same discussion. This to assess whether the metric tends to increase or decrease as a function of pre-existing responses to the same parent post. To evaluate this relationship, we employ the following model specification:

$$y_{i,m} = \alpha_m + \beta_m \, \bar{y}_{i,m}^{\text{sib,older}} + \varepsilon_{i,m}$$

**Table 4 - Linear regression estimates of post-level metrics as a function of mean value of the corresponding metric for older siblings.**

| response metric | term | estimate | std.error | p.value | |
|---|---|---|---|---|---|
| attacking vs respectful | intercept α | -0.04260 | 0.23200 | 0.85500 | |
| | slope β | 0.25200 | 0.04390 | < 0.00001 | *** |
| disagree vs agree | intercept α | -1.40000 | 0.27200 | < 0.00001 | *** |
| | slope β | 0.18400 | 0.05680 | 0.00128 | ** |
| emotion vs fact | intercept α | -0.02420 | 0.22000 | 0.91300 | |
| | slope β | 0.13300 | 0.05720 | 0.02070 | * |

*Sign.level: < 0.001 ~ "***"   < 0.01 ~ "**",  < 0.05 ~ "*",  < 0.1  ~ "†"*

*Note. For each metric, the table reports the intercept and the coefficient, along with cluster-robust standard errors (by discussion), p-values, and significance levels. This regression is based on observations that have at least one older sibling.*

Results from Figure 3 and Table 4 show that a post's tone is strongly shaped by the tone of older siblings within the same conversation subthread. Across all three metrics the coefficient on the mean of older siblings' scores is positive and significant, which means that when earlier sibling posts are more attacking (vs respectful), disagreeing (vs agreeing), or emotional (vs factual), the focal post tends to move in the same direction. The size of these coefficients (ranging from 0.13 to 0.26) indicates a substantial degree of alignment with the local conversation atmosphere, i.e., posts tend to echo the local conversational tone created by earlier posts replying to the same parent post.

*Question: How closely do posts align with the tone and framing of its parent post?*

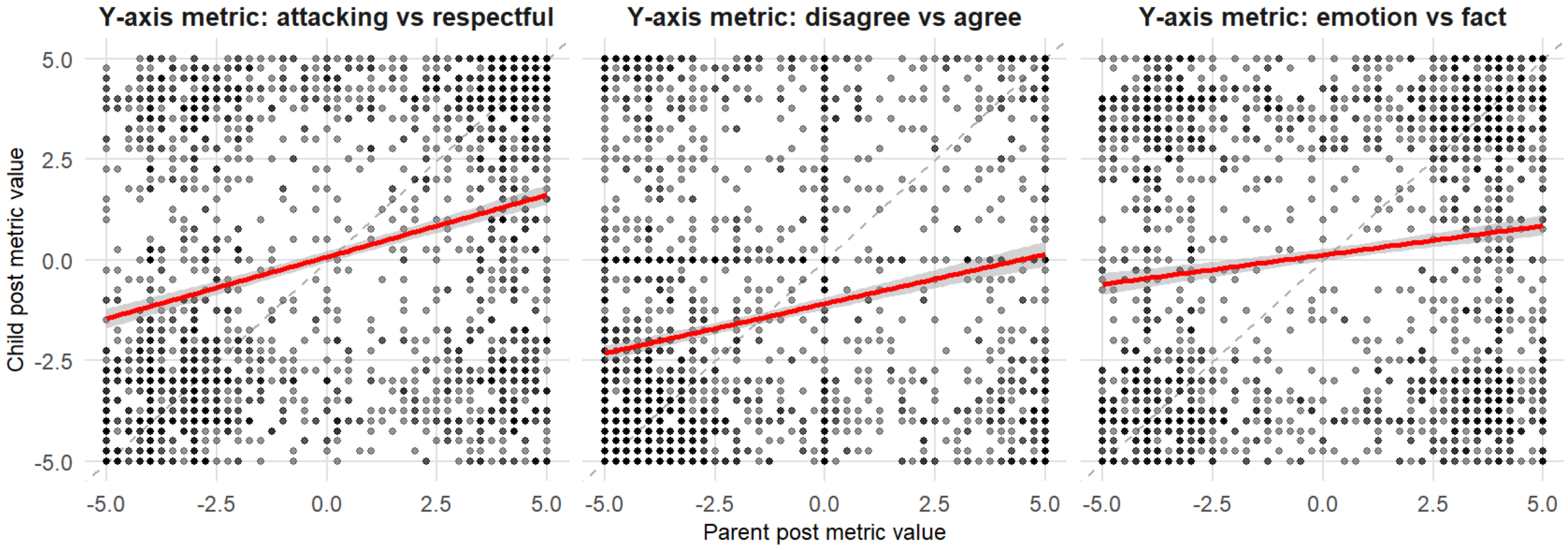

**Figure 4 - Scatter plots of post-level metric values as a function of the parent post metric, with linear regression lines (in red) and associated 95% confidence intervals (in gray).**

To assess whether the metric tends to increase or decrease as a function of the tone of its parent post, for each metric, we estimate a separate linear regression where the value of the metric for a post is related to that of its parent post. Since our metrics are not computed for discussion root posts, all replies to discussion root posts are excluded from this analysis. To evaluate this relationship, we employ the following model specification:

$$y_{i,m} = \alpha_m + \beta_m \, y_{i,m}^{\text{parent}} + \varepsilon_{i,m}$$

**Table 5. -Linear regression estimates of post-level metrics as a function of value of the corresponding metric for parent post.**

| response metric | term | estimate | std.error | p.value | |
|---|---|---|---|---|---|
| attacking vs respectful | intercept α | 0.07780 | 0.14700 | 0.59700 | |
| | slope β | 0.30700 | 0.03020 | < 0.00001 | *** |
| disagree vs agree | intercept α | -1.09000 | 0.13100 | < 0.00001 | *** |
| | slope β | 0.24600 | 0.02670 | < 0.00001 | *** |
| emotion vs fact | intercept α | 0.12600 | 0.16300 | 0.43900 | |
| | slope β | 0.14500 | 0.02810 | < 0.00000 | *** |

*Sign.level: < 0.001 ~ "***"   < 0.01 ~ "**",   < 0.05 ~ "*",   < 0.1 ~ "†"*

*Note. For each metric, the table reports the intercept and the coefficient, along with cluster-robust standard errors (by discussion), p-values, and significance levels. This regression is based on observations for which*

*the parent post metrics are available; that is, all posts excluding replies to the root (first) post of each discussion.*

Results in Figure 4 and Table 5 align with those of Figure 3 and Table 4, showing that a post's tone is strongly shaped by the tone of its parent post. The size of these coefficients (ranging from 0.14 to 0.31) indicates a strong alignment with the parent post.

**Interactions between parent post tone and local conversation atmosphere**

To assess whether there is an interaction between the parent post tone and local conversation atmosphere, for each metric, we estimate a separate linear regression where the value of the metric for a post is jointly related to that of its parent post, that of its older siblings, and to their interaction. Since our metrics are not computed for discussion root posts, all replies to discussion root posts and all posts without at least one older sibling are excluded from this analysis. The specification of this model is the following:

$$y_{i,m} = \alpha_m + \beta_{1,m}\, y_{i,m}^{\text{parent}} + \beta_{2,m}\, \bar{y}_{i,m}^{\text{sib,older}} + \beta_{3,m}\left(y_{i,m}^{\text{parent}} \times \bar{y}_{i,m}^{\text{sib,older}}\right) + \varepsilon_{i,m}$$

**Table 6 - Linear regression estimates of post-level metrics as a function of value of the corresponding metric for parent post, and mean value of the corresponding metric for older siblings.**

| metric | term | estimate | std.error | p.value | |
|---|---|---|---|---|---|
| attacking vs respectful | intercept α | -0.32100 | 0.18600 | 0.08450 | † |
| | slope $\beta_1$ | 0.27300 | 0.04690 | < 0.00001 | *** |
| | slope $\beta_2$ | 0.14700 | 0.03390 | 0.00002 | *** |
| | slope $\beta_3$ | 0.00369 | 0.01630 | 0.82100 | |
| disagree vs agree | intercept α | -1.33000 | 0.27000 | < 0.00001 | *** |
| | slope $\beta_1$ | 0.19300 | 0.05390 | 0.00040 | *** |
| | slope $\beta_2$ | 0.17500 | 0.08590 | 0.04200 | * |
| | slope $\beta_3$ | 0.01790 | 0.01750 | 0.30800 | |
| emotion vs fact | intercept α | -0.19500 | 0.21900 | 0.37300 | |
| | slope $\beta_1$ | 0.23000 | 0.04570 | < 0.00001 | *** |
| | slope $\beta_2$ | 0.09620 | 0.06020 | 0.11100 | |
| | slope $\beta_3$ | 0.03800 | 0.01110 | 0.00071 | *** |

*Sign.level: < 0.001 ~ "***"  < 0.01 ~ "**",  < 0.05 ~ "*",  < 0.1 ~ "†"*

*Note. For each metric, the table reports the intercept and the coefficients, along with cluster-robust standard errors (by discussion), p-values, and significance levels. This regression is based on observations that have at least one older sibling and for which the parent post metrics are available.*

The results presented in Table 6 indicate that both the parent post's metric and the average metric of older siblings play important yet distinct roles in shaping a post's tone and framing across dimensions:

- For attacking vs respectful, both the parent post's value and the average value of older siblings have strong and highly significant positive effects. This suggests that posts tend to align with the tone of both the parent post and the prevailing tone among older siblings, becoming either more attacking (or more respectful) in accordance with these influences. However, the interaction between the parent's tone and older siblings' tone is not significant, indicating that these two factors independently influence a post's alignment within these dimensions.

- For disagree vs agree, the strongly negative intercept confirms an overall baseline tendency toward disagreement in our dataset. However, both the parent post's agreement score and the average agreement score of older siblings have positive and significant coefficients. This indicates that more agreeing parent posts and older siblings independently reduce the overall tendency toward disagreement.

- For emotion vs fact, the parent post's value has a positive and highly significant coefficient, showing that children tend to mirror their parent's tone along the emotional–factual framing spectrum. However, the coefficient of the average older sibling metric is not significant on its own. Interestingly, the interaction between the parent post value and older siblings' mean values is positive and significant, suggesting that alignment with the parent on emotionality or factuality is further strengthened (i.e., multiplicatively amplified) when older siblings also align with the parent on this dimension.

### c. Hypothesis 1.g: Parent–child tone alignment moderation through branch root response stance

*Question: How early branch responses moderate parent–child stance alignment dynamics?*

By interacting the parent post's metric with the sign of the branch root's metric, we can model and test whether branch-root stance conditions the way tone is transmitted from one post to the next. This is important because, as highlighted in the literature review section, online discussions are shaped not only by immediate dyadic interactions, but also by the conversational atmosphere imprinted by the stance polarity of the branch root (i.e., the first reply to the root post that creates a new discussion branch). Reactions to the first post in a discussion (e.g., in terms of agreement vs disagreement) may act as conversation dynamics modulators, establishing norms and expectations that influence subsequent exchanges within that branch.
For example, consider two branches responding to the same discussion root message. In one branch, the first reply expresses strong disagreement with the root post, signaling a confrontational stance. In this context, even moderately critical parent posts may elicit disproportionately negative or conflictual replies, as disagreement has already been normalized within the branch. In contrast, in a branch whose first reply signals agreement with the root post, the same level of parental criticism may be softened, ignored, or reframed constructively.

The model specification that follows allows us to distinguish direct parent–child influence from contextual effects operating at the discussion-branch root level, shedding light on how conflict or consensus emerges and propagates within threaded conversations, from the first reply to deeper levels in that branch.

For the metric *m*, we define the branch-root sign indicator as:

$$\mathrm{BR}_{i,m}^{\mathrm{neg}} = \begin{cases} 1 & \text{if } y_{i,m}^{\text{branch root}} < 0 \\ 0 & \text{otherwise} \end{cases}$$

The specification of this model is the following:

$$y_{i,m} = \alpha_m + \beta_{1,m}\, y_{i,m}^{\text{parent}} + \beta_{2,m}\, \mathrm{BR}_{i,m}^{\text{neg}} + \beta_{3,m}\left(y_{i,m}^{\text{parent}} \times \mathrm{BR}_{i,m}^{\text{neg}}\right) + \varepsilon_{i,m}$$

In relation to this specification, to test Hypothesis 1.g, we are interested in whether the sign of the stance (i.e., agreement vs disagreement) metric at the branch root (indicating initial agreement or disagreement with the root post) acts as a modulating factor for ensuing branch-level conversation dynamics. In particular, to test whether parent–child tone transmission differs systematically between branches that originate in agreement and those that originate in disagreement, thereby capturing how early alignment or opposition shapes the trajectory of interaction within a branch.

**Table 7 - Linear regression estimates of post-level metrics as a function of value of the corresponding metric for parent post, the sign of the branch's root post and the interaction between the two.**

| metric | term | estimate | std.error | p.value | |
|---|---|---|---|---|---|
| disagree vs agree | intercept α | -0.92099 | 0.11064 | <0.00001 | *** |
| | slope $\beta_1$ | 0.32972 | 0.02940 | <0.00001 | *** |
| | slope $\beta_2$ | -0.39631 | 0.16307 | 0.0152 | * |
| | slope $\beta_3$ | -0.19115 | 0.04381 | 0.00001 | *** |

*Sign.level: < 0.001 ~ "***"   < 0.01 ~ "**",  < 0.05 ~ "*",  < 0.1  ~ "†"*
*Note. For each metric, the table reports the intercept and the coefficients, along with cluster-robust standard errors (by discussion), p-values, and significance levels. This regression is based on observations that have at least one older sibling and for which the parent post metrics are available.*

The negative and highly significant intercept shown in Table 7 indicates a baseline tendency toward disagreement when both the parent post's agreement score and the branch-root indicator are at zero. Consistent with earlier results, the coefficient on the parent post's agreement score is positive and highly significant, indicating strong parent–child alignment: replies tend to become more agreeing as the parent post becomes more agreeing. The coefficient on BR, the branch-root (-) sign indicator, is negative and statistically significant, suggesting that, holding other factors constant, branches initiated by disagreement with the discussion root exhibit a stronger baseline tendency toward disagreement. This finding is consistent with the idea that early branch-level responses establish local norms that shape subsequent interaction.
The interaction term between the parent post's agreement score and BR is also negative and highly significant, indicating that parent–child agreement alignment is weaker in branches whose first reply expresses disagreement with the root post, and stronger in branches whose first reply expresses agreement. In other words, while replies generally mirror the stance of their immediate parent, this alignment intensity is moderated by the tone established at the start of the branch.

## Conclusions

This study examined how conflictual language emerges and propagates within online threaded conversations, such as those of Forums and Social Media like Reddit, by focusing on the temporal and structural geometry of interaction. Using 4forums.com Climate Change–related discussions from the Internet Argument Corpus and LLM-based annotations of stance and tone, we showed that conflict is not merely a property of message content, but a dynamic process shaped by conversational timing, local alignment, and early interactional signals embedded in discussion graphs.

First, our results demonstrate that conversational timing indeed matters, but in systematic and nuanced ways: Longer delays between successive posts are associated with more respectful replies, suggesting that slower conversational rhythms may attenuate impulsive or affectively charged reactions. At the same time, delays relative to the parent post are linked to weaker disagreement and increased emotional framing, indicating that temporal distance from the triggering contribution can alter both stance and affective expression. These findings refine existing work on conversational rhythm by showing that different temporal reference points—previous post versus parent post—capture distinct mechanisms in the evolution of conflictual discourse in threaded discussions.

Second, we find strong evidence of new post alignment with the local conversational environment. Replies consistently mirror the tone and stance of both their parent post and earlier sibling replies, with parent effects generally stronger than sibling effects. This alignment holds across dimensions of respectfulness, agreement, and emotional versus factual framing, confirming that local conversational context exerts a powerful influence on how users express disagreement and civility. Importantly, these effects are largely additive for civility-related dimensions, suggesting independent channels through which parent and sibling contexts shape subsequent contributions.

Third, our analysis reveals that early branch-level responses play an important moderating role in the propagation of conflict. Branches initiated by disagreement with the discussion root exhibit stronger baseline disagreement and weaker parent–child alignment, whereas branches initiated by agreement amplify alignment effects. This finding supports the idea that early interactional signals establish local norms that condition how tone and stance are transmitted as discussions unfold, highlighting the disproportionate influence of the first replies to the root post (i.e., first message in a discussion) in shaping the trajectory of a conversational branch.

Taken together, these results advance a discussion-graph-based perspective on conflictual discourse, demonstrating the value of modeling online conversations as structured, temporally evolving processes rather than as collections of independent messages. Beyond their theoretical implications, our findings suggest practical avenues for intervention: platform designs or moderation tools that attend to conversational timing, early branch dynamics, and local tone alignment may help mitigate the escalation of conflict without suppressing disagreement. More broadly, by identifying the micro-level interactional mechanisms through which conflict propagates, this work contributes to a deeper understanding of how online discourse shapes the quality of democratic debate in polarized and science-intensive domains.

Regarding the main limitations of this work and directions for future research, we note that our annotations are currently based on a single commercial LLM model (ChatGPT-5.1-mini) and may therefore be subject to model-specific biases related to the training set, architecture, guardrails and alignment procedures. Future work could address this limitation by incorporating multiple models or human–LLM hybrid annotation schemes. In terms of modeling, we do not analyze lagged interdependencies among the different annotated dimensions, nor do we examine how conflictual language emerges as a byproduct of longer conversational histories or repeated interactions between the same user pairs across discussions, where trust, familiarity, or antagonism may accumulate over time. These limitations point to several promising directions for future work, including the characterization of users by conversational conflict styles based on systematic deviations in tone stance and emotional framing between their replies and the parent posts they respond to, an approach we plan to pursue in subsequent research.

## References

Abbott, R., Ecker, B., Anand, P., & Walker, M. A. (2016). *Internet Argument Corpus 2.0: An SQL schema for dialogic social media and the corpora to go with it.* In N. Calzolari, K. Choukri, T. Declerck, S. Goggi, M. Grobelnik, B. Maegaard, J. Mariani, H. Mazo, A. Moreno, J. Odijk, & S. Piperidis (Eds.), *Proceedings of the Tenth International Conference on Language Resources and Evaluation (LREC 2016)* (pp. 4445–4452). European Language Resources Association (ELRA). https://aclanthology.org/L16-1704/

Altenburger, K. M., Kraut, R. E., Hayati, S. A., Dwivedi-Yu, J., Peng, K., & Wang, Y.-C. (2024). Consequences of Conflicts in Online Conversations. *Proceedings of the International AAAI Conference on Web and Social Media, 18*, 57–69. https://doi.org/10.1609/icwsm.v18i1.31297

Anjum, & Katarya, R. (2024). Hate speech, toxicity detection in online social media: A recent survey of state of the art and opportunities. *International Journal of Information Security,* 23(1), 577–608. https://doi.org/10.1007/s10207-023-00755-2

Baughan, A., Petelka, J., Yoo, C. J., Lo, J., Wang, S., Paramasivam, A., Zhou, A., & Hiniker, A. (2021). Someone Is Wrong on the Internet: Having Hard Conversations in Online Spaces. *Proceedings of the ACM on Human-Computer Interaction*, 5(CSCW1), 1–22. https://doi.org/10.1145/3449230

Bennett, W., Pfetsch, B., Rethinking Political Communication in a Time of Disrupted Public Spheres, *Journal of Communication*, Volume 68, Issue 2, April 2018, Pages 243–253, https://doi.org/10.1093/joc/jqx017

Böhm, G., & Pfister, H.-R. (2025). Exploring climate change discourses on the internet: A topic modeling study across ten years. *Journal of Risk Research*, 28(3–4), 203–230. https://doi.org/10.1080/13669877.2024.2387337

Bolander, B., & Locher, M. A. (2017). 22. Conflictual and consensual disagreement. In C. Hoffmann & W. Bublitz (Eds.), *Pragmatics of Social Media* (pp. 607–632). De Gruyter. https://doi.org/10.1515/9783110431070-022

Brown, P., & Levinson, S. C. (1987). *Politeness: Some universals in language usage*. Cambridge University Press

Canute, M., Jin, M., Holtzclaw, H., Lusoli, A., Adams, P., Pandya, M., Taboada, M., Maynard, D., & Chun, W. H. K. (2023). Dimensions of Online Conflict: Towards Modeling Agonism. *Findings of the Association for Computational Linguistics: EMNLP 2023*, 12194–12209. https://doi.org/10.18653/v1/2023.findings-emnlp.816

Choi, Y., & Cardie, C. (2008). Learning with Compositional Semantics as Structural Inference for Subsentential Sentiment Analysis. *EMNLP 2008,* 793–801. https://aclanthology.org/D08-1083.pdf

Danescu-Niculescu-Mizil, C., Sudhof, M., Jurafsky, D., Leskovec, J., & Potts, C. (2013). A computational approach to politeness with application to social factors. *Proceedings of the 51st Annual Meeting of the Association for Computational Linguistics* (ACL), Volume 1 (Long Papers). https://arxiv.org/pdf/1306.6078

D'Costa, P. R., Rowbotham, E., & Hu, X. E. (2024). What you say or how you say it? Predicting Conflict Outcomes in Real and LLM-Generated Conversations (arXiv:2409.09338). arXiv. https://doi.org/10.48550/arXiv.2409.09338

Ferrara, E., & Yang, Z. (2015). Measuring emotional contagion in social media. *PLoS ONE*, 10(11), e0142390. https://doi.org/10.1371/journal.pone.0142390

Habernal, I., Wachsmuth, H., Gurevych, I., & Stein, B. (2018). Before Name-Calling: Dynamics and Triggers of Ad Hominem Fallacies in Web Argumentation. *Proceedings of the 2018 Conference of the North American Chapter of the Association for Computational Linguistics: Human Language Technologies*, Volume 1 (Long Papers), 386–396. https://doi.org/10.18653/v1/N18-1036

Imran, M. M., Zita, R., Rahman, R. R., Chatterjee, P., & Damevski, K. (2025). Toxicity Ahead: Forecasting Conversational Derailment on GitHub (arXiv:2512.15031). arXiv. https://doi.org/10.48550/arXiv.2512.15031

Jahan, M. S., & Oussalah, M. (2023). A systematic review of hate speech automatic detection using natural language processing. *Neurocomputing*, 546, 126232. https://doi.org/10.1016/j.neucom.2023.126232

Korre, K., Tsirmpas, D., Gkoumas, N., Cabalé, E., Myrtzani, D., Evgeniou, T., Androutsopoulos, I., & Pavlopoulos, J. (2025). Evaluation and Facilitation of Online Discussions in the LLM Era: A Survey. *Proceedings of the 2025 Conference on Empirical Methods in Natural Language Processing*, 24454–24473. https://doi.org/10.18653/v1/2025.emnlp-main.1243

Leippold, M., Vaghefi, S. A., Stammbach, D., Muccione, V., Bingler, J., Ni, J., ... & Huggel, C. (2025). Automated fact-checking of climate claims with large language models. *npj Climate Action*, 4(1), 17. https://doi.org/10.1038/s44168-025-00215-8

Le Mens, G., & Gallego, A. (2024). Positioning political texts with large language models by asking and averaging. *Political Analysis.* https://doi.org/10.1017/pan.2024.29

Liu, Y., Kliman-Silver, C., & Mislove, A. (2014). The Tweets They Are a-Changin': Evolution of Twitter Users and Behavior. *Proceedings of the International AAAI Conference on Web and Social Media*, 8(1), 305–314. https://doi.org/10.1609/icwsm.v8i1.14508

Ma, S., Chen, Q., Wang, X., Zheng, C., Peng, Z., Yin, M., & Ma, X. (2025, April). Towards human-ai deliberation: Design and evaluation of llm-empowered deliberative ai for ai-assisted decision-making. In *Proceedings of the 2025 CHI Conference on Human Factors in Computing Systems* (pp. 1-23). https://doi.org/10.1145/3706598.371342

Myers, T. A., Roser-Renouf, C., & Maibach, E. (2023). Emotional responses to climate change information and their effects on policy support. *Frontiers in Climate*, 5, 1135450. https://doi.org/10.3389/fclim.2023.1135450

Ranjith, S., Chowdary, C. R., & Tiwari, P. (2025). Learning models to forecast toxicity in conversation threads by identifying potential toxic users. *Evolving Systems*, 16(1), 8. https://doi.org/10.1007/s12530-024-09639-9

Roos, C. A., Koudenburg, N., & Postmes, T. (2020). Online Social Regulation: When Everyday Diplomatic Skills for Harmonious Disagreement Break Down. *Journal of Computer-Mediated Communication*, 25(6), 382–401. https://doi.org/10.1093/jcmc/zmaa011

Somasundaran, S., Namata, G., Wiebe, J., & Getoor, L. (2009). Supervised and Unsupervised Methods in Employing Discourse Relations for Improving Opinion Polarity Classification. Proceedings of the 2009 *Conference on Empirical Methods in Natural Language Processing* (pp.170–179). https://aclanthology.org/D09-1018.pdf

Somasundaran, S., & Wiebe, J. (2010, June). Recognizing stances in ideological on-line debates. In *Proceedings of the NAACL HLT 2010 workshop on computational approaches to analysis and generation of emotion in text* (pp. 116-124). https://dl.acm.org/doi/10.5555/1860631.1860645

Tan, Z., Li, D., Wang, S., Beigi, A., Jiang, B., Bhattacharjee, A., Karami, M., Li, J., Cheng, L., & Liu, H. (2024). Large language models for data annotation and synthesis: A survey. In *Proceedings of the 2024 Conference on Empirical Methods in Natural Language Processing (EMNLP)* (pp. 930–957). Association for Computational Linguistics.

Tsvetkova, M., García-Gavilanes, R., & Yasseri, T. (2016). Dynamics of Disagreement: Large-Scale Temporal Network Analysis Reveals Negative Interactions in Online Collaboration. *Scientific Reports*, 6(1), 36333. https://doi.org/10.1038/srep36333

Tyagi, A., Uyheng, J., & Carley, K. M. (2021). Heated conversations in a warming world: Affective polarization in online climate change discourse follows real-world climate anomalies. *Social Network Analysis and Mining*, 11(1), 87. https://doi.org/10.1007/s13278-021-00792-6

Vaswani, A., Shazeer, N., Parmar, N., Uszkoreit, J., Jones, L., Gomez, A. N., Kaiser, Ł., & Polosukhin, I. (2017). Attention is all you need. *Advances in Neural Information Processing Systems (NeurIPS 2017).* (arXiv:1706.03762).

Walker, M. A., Fox Tree, J. E., Anand, P., Abbott, R., & King, J. (2012). A corpus for research on deliberation and debate. In N. Calzolari, K. Choukri, T. Declerck, M. Uğur Doğan, B. Maegaard, J. Mariani, A. Moreno, J. Odijk, & S. Piperidis (Eds.), *Proceedings of the Eighth International Conference on Language Resources and Evaluation* (LREC 2012) (pp. 812–817). European Language Resources Association (ELRA).

Wang, L., & Cardie, C. (2016). A Piece of My Mind: A Sentiment Analysis Approach for Online Dispute Detection. https://arxiv.org/pdf/1606.05704v1

Zhang, J., Chang, J., Danescu-Niculescu-Mizil, C., Dixon, L., Hua, Y., Taraborelli, D., & Thain, N. (2018). Conversations Gone Awry: Detecting Early Signs of Conversational Failure. *Proceedings of the 56th Annual Meeting of the Association for Computational Linguistics*, Volume 1 (Long Papers), 1350–1361. https://doi.org/10.18653/v1/P18-1125

## Acknowledgements & Funding

The authors thank Arwa Rahmani for excellent support in the project, in particular for identifying, preparing and describing the dataset, developing scripts to submit data for annotation to the OpenAI APIs, testing the LLM prompts, and running the LLM-based annotations.

The authors also wish to thank Massimiliano Nuccio for preliminary discussions on LLM based annotations, Andrea Baronchelli and the participants in the Venice Summer Workshop for fruitful discussions and feedback on an earlier version of the project.

The authors (C.S., C.C.) acknowledge funding from the European Union's Horizon Europe programme under grant agreement ID 101094752: Social Media for Democracy (SoMe4Dem).

C.S. acknowledges funding from Inria.

## Competing Interests

The authors declare that they have no competing interests related to this work. There are no financial, personal, or professional relationships that could be perceived as influencing this research or its findings.

## Authors' Contribution Statement

Conceptualization: all (C.S., C.C.);
Methodology: all (C.S., C.C.);
Formal Analysis: C.S.;
Investigation: all (C.S., C.C.);
Data Curation: C.S. (with the support of the Arwa Rahmani);
Software: C.S.;
Visualization: C.S.;
Writing – Original Draft Preparation: all (C.S., C.C.);
Writing – Review & Editing: all (C.S., C.C.);
Funding Acquisition : (C.C.);

## Usage of generative AI

The authors and Arwa Rahmani used OpenAI's ChatGPT-5.1-mini model to generate post-level annotations as described in the Methods and Data section. All interactions with the model were conducted via the OpenAI API in accordance with OpenAI's privacy and data protection policies, which state that API data are not used for model training and are handled in compliance with applicable data protection regulations, including the EU General Data Protection Regulation (GDPR).
Relevant documentation is available at:

- https://openai.com/security-and-privacy/
- https://openai.com/business-data/

.

The Educational edition of GitHub Copilot Pro (Claude and ChatGPT models) has been used for proofreading some of the sections of this work and for minor editing assistance. GitHub Copilot was used in a mode in which text and data is not retained or used for model training.